\documentclass[12pt,english,twoside]{article}

\usepackage[numbers]{natbib}
\usepackage{url}
\makeatletter
\def\url@leostyle{%
    \def\UrlFont{\sf}}{\def\UrlFont{\small\ttfamily}}
\makeatother
\usepackage{geometry}
\usepackage{fancyhdr}
\usepackage{pstricks}

\usepackage[font=small,labelfont=bf]{caption}

\usepackage[hidelinks,breaklinks]{hyperref}
\usepackage[hyphenbreaks]{breakurl}

\usepackage{authblk}

\usepackage{amsmath}

\usepackage{amssymb}

\usepackage{nicefrac}

\newcommand{\ignore}[1]{}

\usepackage{color}

\usepackage{color}

\usepackage[T1]{fontenc}
\usepackage{charter} 

\usepackage{microtype}

\usepackage{marginnote}

\usepackage{graphicx}

\usepackage[english]{babel}

\numberwithin{equation}{section}

\newcommand{\shortTitle}{The Measurement Problem Is a Feature, Not a Bug}
\newcommand{\documentTitle}{\shortTitle~--~Schematising the Observer and the Concept of an Open System on an Informational, or (neo-)Bohrian, Approach\thanks{This paper has been published in {\color{blue} \href{https://doi.org/10.3390/e25101410}{\emph{Entropy} 2023, 25(10), 1410}}. Because \emph{Entropy} has an editorial policy against footnotes, this version, which uses footnotes but is otherwise identical to the published version, may be more readable. Please refer to the publication details just mentioned when citing this paper. Note that the page numbers (indicated in the margins) below are the same as in the published version.}}

\title{\documentTitle}
\author{Michael E. Cuffaro\footnote{Munich Center for Mathematical Philosophy,
      Ludwig-Maximilians-Universit\"at M\"unchen,
      E-mail: Michael.Cuffaro@lmu.de}}

\begin{document}

\maketitle

\thispagestyle{empty}

\begin{abstract}I flesh out the sense in which the informational approach to interpreting quantum mechanics, as defended by Pitowsky and Bub and lately by a number of other authors, is (neo-)Bohrian. I argue that on this approach, quantum mechanics represents what Bohr called a ``natural generalisation of the ordinary causal description'' in the sense that the idea (which philosophers of science like Stein have argued for on the grounds of practical and epistemic necessity) that understanding a theory as a theory of physics requires that one be able to ``schematise the observer'' within it is elevated in quantum mechanics to the level of a postulate in the sense that interpreting the outcome of a measurement interaction, as providing us with information about the world, requires as a matter of principle, the specification of a schematic representation of an observer in the form of a `Boolean frame'---the Boolean algebra representing the yes-or-no questions associated with a given observable representative of a given experimental context. I argue that the approach's central concern is with the methodological question of how to assign physical properties to what one takes to be a system in a given experimental context, rather than the metaphysical question of what a given state vector represents independently of any context, and I show how the quantum generalisation of the concept of an open system may be used to assuage Einstein's complaint that the orthodox approach to quantum mechanics runs afoul of the supposedly fundamental methodological requirement to the effect that one must always be able, according to Einstein, to treat spatially separated systems as isolated from one another.

\end{abstract}


\section{Introduction}
\label{s:introduction}

Niels Bohr's views on quantum mechanics, and on the methodology of physics more generally \citep{demopoulosOnTheories, perovic2021}, have been the subject of renewed attention in recent years, both in the foundations and philosophy of physics as well as in the area of general philosophy of science (see, for instance, \citep{evans2020}). In the former area, this has taken the form of a number of approaches to interpreting the formalism that in some sense claim to be a modern expression of Bohr's approach (see, for instance, \citep{brukner2017} (p.\ 98); \citep{bub2017}; \citep{demopoulosOnTheories} (chp.\ 4); \citep{3m2020} (p.\ 16); \citep{landsman2017} \mbox{(p.\ viii)}), as well as approaches that take themselves to be inspired by certain aspects of Bohr's approach, but that otherwise depart from it in various ways (see, for instance, \citep{fuchs2017} (sct.\ 1); \citep{healey2017} (pp.\ 253--254); \citep{rovelli2021} (pp.\ 135--142)).

The words `informational' or `information-theoretic' are, or have been, used to describe many of the approaches in both categories. But my focus in this paper will specifically be on the informational approach that one can trace to the work of Itamar Pitowsky \citep{pitowsky1989}, later further developed in conjunction with Jeffrey Bub \citep{bub-pitowsky2010, bub2016, bub2017, bub2018a, bubRedux, bub2021} and others. The most recent book-length elaboration and defence of the approach (which is what I will mainly be drawing on here) is by \citet*[]{3m2020}, which also draws on the closely related ideas of William Demopoulos \citep{demopoulosOnTheories}.\footnote{Note that although it was only published in 2022, the same year as \citep{3m2020}, Demopoulos's book, On Theories, was completed shortly before his death in 2017.} \marginnote{\textbf{2}} Although some of the concepts and arguments have been adapted and clarified over the years, the core of the view developed in \citep{3m2020} and, independently, in Bub's later work remains unchanged from the one defended by \citet[]{bub-pitowsky2010}.\footnote{The book by Janas, Cuffaro \& Janssen will be my primary source for the exposition which follows, but note that I will occasionally refer to earlier works as well to call attention to some of the connections between them. Note also that although there is no difference of substance between \citep{3m2020} and these other works, there are of course differences in what is emphasised in each of these works and in the particular issues each deals with.}

Defenders of this informational approach to the interpretation of quantum mechanics think of their view as (neo-)Bohrian in the sense of amounting to a rehabilitation of Bohr---or at least what they take to be essential to Bohr's view---and my aim in this paper will be to flesh this out. The reader should keep in mind, however, that it is not one of the goals of the research programme, per se, to contribute to the historical scholarship on Bohr, and it will not be my goal here. The upshot is that one may (if one is so inclined) call the approach that we will be discussing Bohrian if one agrees that it has correctly characterised the historical views of Bohr. Otherwise, one may call it neo-Bohrian. Such labels are ultimately not \mbox{my concern.}\footnote{Note that for their views on Bohr, I am drawing on \citep{bub2017, cuffaro2010, cuffaro2018d, cuffaroPerovicReview, demopoulosOnTheories}, as well as long-running personal communication with Jeffrey Bub, (the late) William Demopoulos and Michel Janssen.}

As for the rest of this paper, it will be framed in terms of the following passage that one can find in a letter dated 24 March 1928 that Bohr sent to Paul Dirac. Note that, below, ``my article'' refers to what later would come to be known as the `Como paper' \citep{bohr1928}, wherein Bohr had just laid out his considered views on the (then) new quantum mechanics.\footnote{The paper was based on a lecture Bohr gave at a conference in Como in northern Italy in 1927, honouring the centenary of Alessandro Volta's death. For more on the period leading up to its publication, see \citep{degregorio2014}.}

\begin{quote}
  I quite appreciate your remarks that in dealing with observations we always witness through some permanent effects a choice of nature between the different possibilities. However, it appears to me that the permanency of results of measurements is inherent in the very idea of observation; whether we have to do with marks on a photographic plate or with direct sensations the possibility of some kind of remembrance is of course the necessary condition for making any use of observational results. It appears to me that the permanency of such results is the very essence of the ordinary causal space-time description. This seems to me so clear that I have not made a special point of it in my article. \emph{What has been in my mind above all [, rather,] was the endeavour to represent the statistical quantum theoretical description as a natural generalisation of the ordinary causal description} and to analyze the reasons why such phrases like a choice of nature present themselves in the description of the actual situation. In this respect it appears to me that the emphasis on the subjective character of the idea of observation is essential. Indeed I believe that the contrast between this idea and the classical idea of isolated objects is decisive for the limitation which characterises the use of all classical concepts in the quantum theory. Especially in relation with the transformation theory the situation may, I think, be described by saying that any such concepts can be used unaltered if only due regard is taken to the unavoidable feature of complementarity. \citep{bohrToDirac1928} (pp.\ 45--46, emphasis added).
\end{quote}

In the sequel, I will unpack this (in the conceptual, not historical, sense), or at any rate, what those who advocate for the informational approach we are discussing here, take Bohr to be conveying to Dirac in this passage. I will begin, in Section \ref{sec:necessary_conditions} (entitled ``the necessary conditions for making any use of observational results''), by discussing what the mathematical logician, George Boole, described as the `conditions of possible experience' in relation to the observation of statistical data. I then use Pitowsky's work on correlation polytopes to motivate a particular setup involving three correlated random variables for which \citet*[]{3m2020} provide a \marginnote{\textbf{3}} visual representation of the part of the space of possible correlations between the variables that can be represented in a local hidden-variable theory and in quantum mechanics, respectively. In Section \ref{sec:natural_generalisation} (entitled ``quantum mechanics as a natural generalisation of ordinary causal description''), I explain the sense in which one can understand quantum mechanics to be a generalisation of the kind of description that classical mechanics makes precise. In particular, I explain (in Section \ref{sec:new_kinematics}) that the significant differences between quantum and classical mechanics are traceable to the constraints each conceptual framework imposes on our representations of systems independently of the specifics of their dynamics. In the classical case, these constraints allow for a globally Boolean description of what one naturally thinks of as the properties of a system. In the quantum case, they do not. Howard \citet[]{stein1994} and Erik \citet[]{curiel2020} have argued that understanding a theory, as a theory of physics, requires that one ``schematise the observer'' within it in a sense that goes beyond the pure formalism that one uses to characterise a system. The basic idea is that we do not know how, given our current epistemic state in relation to our best theories of physics, to understand those theories \emph{as} theories of physics without understanding how to schematise the observer in that sense.\footnote{Note that Curiel (personal communication) does not deny that this circumstance could change in the future, though he would argue that it is difficult to imagine (at least in our current epistemic state) what such a change would look like.}

The advocate of the informational, or (neo-)Bohrian, approach agrees, and in Section \ref{sec:subjective_character}, I draw on Bohr to help illuminate why. I argue that what Stein and Curiel have, on the grounds of practical and epistemic necessity, claimed to be required in classical theory should be understood, for a (neo-)Bohrian, to be elevated within quantum theory to the level of a postulate, in the sense that interpreting the outcome of a measurement interaction as providing us with information about the world requires, as a matter of principle, the specification of a schematic representation of an observer in the form of a `Boolean frame'---the Boolean algebra representing the yes-or-no questions associated with a given observable representative of a given experimental context. In Section \ref{sec:isolated_objects}, I discuss the quantum generalisation of the concept of an open system that is implied by this, and how it helps to assuage one of Einstein's complaints, against the orthodox interpretation of quantum mechanics, that it runs afoul of a fundamental methodological requirement to the effect that one must always be able, according to Einstein, to treat spatially separated systems as isolated from \mbox{one another.}

Note that I am not necessarily claiming that defenders of the approach to quantum mechanics under discussion would endorse everything that Stein (or Curiel) has to say about quantum mechanics (and vice versa) even though I am convinced that there is a substantial amount that they do agree on. I disagree, in particular, with Stein's comment to the effect that: ``In [quantum mechanics] we just do not know how to `schematize' the observer and the observation'' \citep{stein1994} (p.\ 653) while agreeing with him that, at least on the approach under discussion here, ``the difficulties [quantum mechanics] presents arise from the fact that \emph{the mode in which this theory `represents' phenomena} is a radically novel one'' (\citep{stein1989} (p.\ 59), emphasis is original; cited in \citep{stein1994} (p.\ 653)). (For Stein's views on quantum mechanics, see \citep{stein1972}.)

Regarding the question of what a given Boolean frame represents physically, the answer, in short, is that it represents anything that can be used to instantiate it using the available means (see \citep[]{3m2020} (pp.\ 202--213); cf. \citep[]{myrvold2011}). It must not be forgotten, however, that on the approach we are discussing there is only ever \emph{one} experimental context represented by a Boolean frame that is associated with any given observable, which is left out of the quantum description as a matter of principle.


\section{The Necessary Conditions for Making Any Use of Observational Results}
\label{sec:necessary_conditions}

Many things are meant by `phenomena' and by `observation', but in the context of physics, the relevant aspect of both that concerns us here is that they can be mathematically represented \citep{bogenwoodward1988}. Newton, famously, appealed to the phenomenon that ``[t]he circumjovial \marginnote{\textbf{4}} planets, by radii drawn to the center of Jupiter, describe areas proportional to the times, and their periodic times---the fixed stars being at rest---are as the $\nicefrac{3}{2}$ powers of their distances from that center'' \citep{newton1999} (p.\ 797). On the basis of it and similarly mathematised observations concerning the motions of the circumsaturnian planets (i.e., the moons of Saturn) and those of the five so-called primary planets: Mercury, Venus, Mars, Jupiter and Saturn---such as the phenomenon that ``[t]he periodic times of the five primary planets and of either the sun about the earth or the earth about the sun---the fixed stars being at rest---are as the $\nicefrac{3}{2}$ powers of their mean distances from the sun'' \citep{newton1999} (p.\ 801)---Newton argued to the conclusion that there is a force called gravity through which every material object in the universe is attracted, to a certain degree, to every other \citep{harper2011, smith2002}.

\begin{figure}[h]
  \begin{center}
    \mbox{} \\[12pt]
    \begin{tabular}{l l}
      \begin{tabular}{l | l | l | l | l}
        $E_1$ & $E_2$ & \dots   & $E_n$ \\ \hline
        0   & 0  & \dots   & 1 \\ \hline
        0   & 1  & \dots   & 0 \\ \hline
        \vdots & \vdots & \vdots & \vdots
      \end{tabular}
      &
      \qquad\qquad
      \raisebox{-30pt}{
        \begin{pspicture}(1,1,1)
          \psset{linecolor=black,unit=3}
          \psline[](0,0,0)(1,1,1)
          \psline[](0,0,0)(0.35,0.65,0)
          \psline[](0.35,0.65,0)(1,1,1)
          \psline[](1,1,1)(0.5,0,0)
          \psline[](0,0,0)(0.5,0,0)
          \psline[linestyle=dotted](0.35,0.65,0)(0.5,0,0)
        \end{pspicture}
      }
    \end{tabular}
  \end{center}
  \caption{At the (\textbf{left}): A table describing consistent assignments of extremal probabilities to a set of logically connected events $E_1$, $\dots$, $E_n$, convex combinations of which can be visualised as a polytope, the facets of which (like those of the three-dimensional polytope depicted on the (\textbf{right})) are associated with linear inequalities of the form $a_1p_1 + a_2p_2 + \dots + a_np_n + a \geq 0$, where the $p_i$ are extremal probabilities. One of these is a Bell inequality.\\}
  \label{fig:algorithm}
\end{figure}
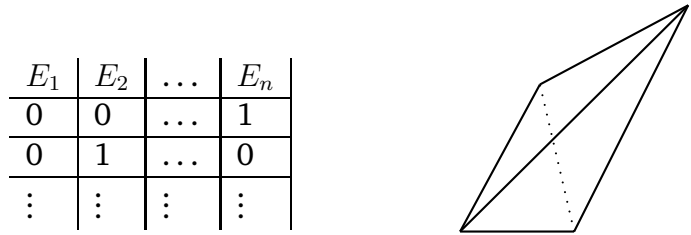

In the 19th century, the philosopher and mathematical logician, George Boole, described a number of what he called `conditions of possible experience' in relation to the observation of statistical data that, he argued, are such that ``[w]hen satisfied they indicate that the data \emph{may} have, when not satisfied they indicate that the data \emph{cannot} have resulted from an actual observation.'' (\citep{boole1862} (p.\ 229), cited in \citep{pitowsky1994} (p.\ 100)). Boole explicates the concept in the following way (the notation is Pitowsky's): given the rational numbers $p_1$, $\dots$, $p_n$ representing the relative frequencies of $n$ in general logically connected events $E_1$, $\dots$, $E_n$ (Pitowsky \citep{pitowsky1994} (p.\ 2) gives the following example: E1: it will rain in Paris tomorrow; E2: it will rain in Madrid tomorrow; E3: it will rain in Paris and in Madrid tomorrow), the \emph{conditions of possible experience} with respect to that data are the necessary and sufficient conditions under which the $p_i$ can be realised as probabilities corresponding to the $E_i$ in some probability space. They are yielded by the following algorithm: begin by writing down a `truth table' (see Figure \ref{fig:algorithm}) whose rows are the vectors, $(p_1$, $\dots$, $p_n)$, describing the consistent assignments (given their logical connections) of extremal probabilities to $E_1$, $\dots$, $E_n$. Now take the convex hull of these vectors. This yields a polytope, the facets of which are associated with a number of linear inequalities, special cases of which include the one associated with John S. Bell and its variants \citep{pitowsky1994} (pp.\ 103--104).\\

\begin{figure}[h]
   \includegraphics[width=3.5in]{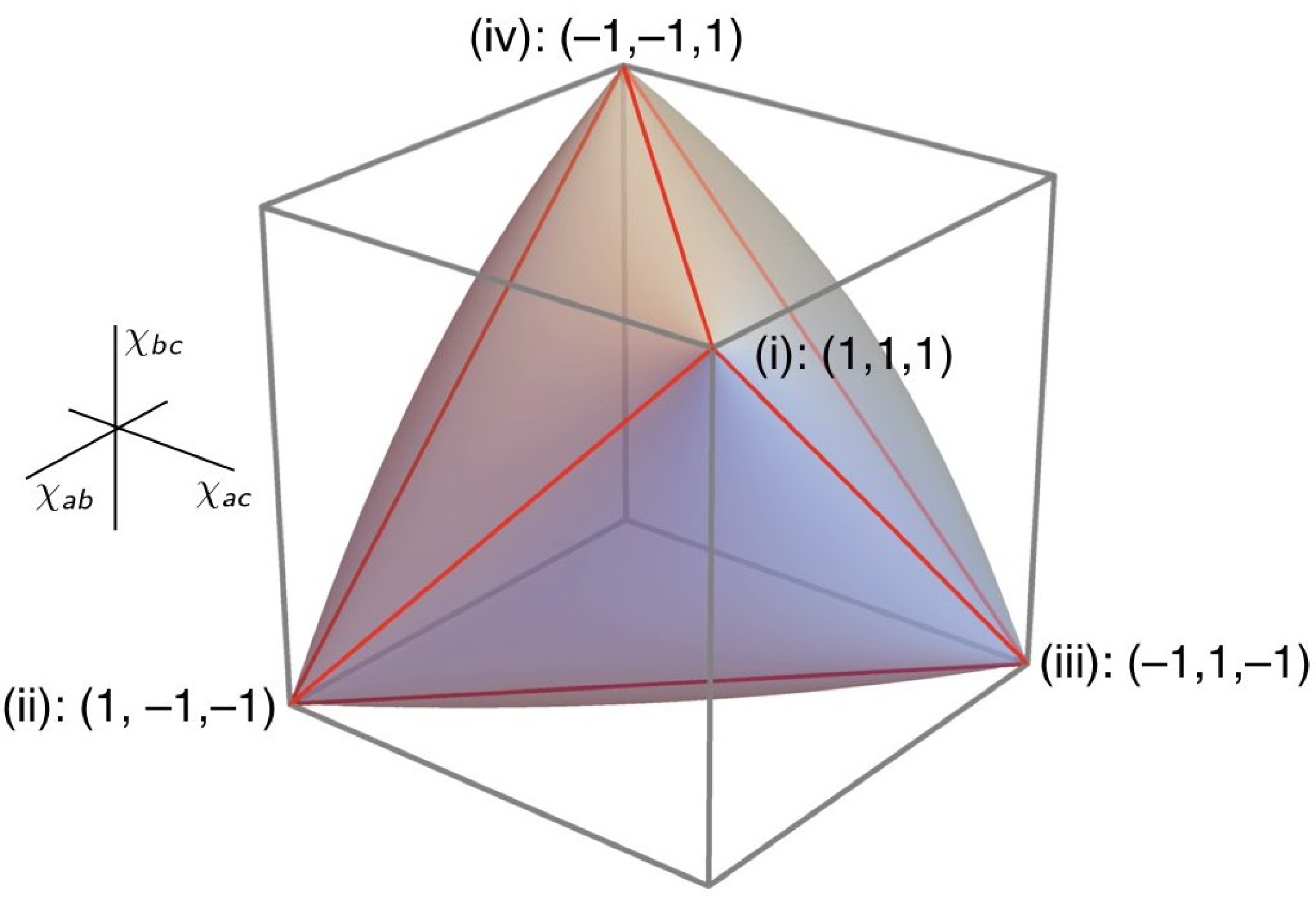} 
   \caption{Elliptope of triplets of anti-correlation coefficients $(\chi_{ab}, \chi_{ac}, \chi_{bc})$ representing the allowed correlations, according to quantum mechanics, in the Mermin-inspired setup considered in \citep{3m2020}. Note that the anti-correlation coefficient, $\chi_{ab}$, which parametrisess the correlations between the values obtained by the two distinct parties in the Mermin-inspired setup, is just the negative of its corresponding Pearson correlation coefficient: $\rho_{XY}$ \citep{3m2020} (p.\ 72--73). Image source: \citep{3m2020} (p.\ 46).\\}
   \label{fig:elliptope}
\end{figure}

In the spirit of Pitowsky's further work on correlation polytopes \citep{pitowsky1989, pitowsky1991, pitowsky2008}, and building on Bub's work on correlation arrays \citep{bub2016}, \citet*[]{3m2020} consider a particular setup inspired by David \citet[]{mermin1981} in which two parties are given one of two correlated systems each and are each asked to measure their system using one of three possible settings. In the setup considered in \citep{3m2020}, the possible outcomes of the three types of measurements are represented by the balanced random variables $X$, $Y$ and $Z$ such that a possible value $x$ of $X$ can be an element of a discrete set $\{ x_i \}$ or a continuous interval $[a, b]$ of real numbers or an element of the union of such sets and intervals, and a random variable $X$ is called \emph{balanced} if and only if: (1) whenever $x$ is a possible value, $\mathrm{-}x$ is a possible value; and (2) $x$ is as likely as $\mathrm{-x}$, i.e., $\mathrm{Pr}(x) = \mathrm{Pr}(\mathrm{-}x)$ (see \citep{3m2020} (pp.\ 67--68)). There is a nonlinear \marginnote{\textbf{5}} constraint on the correlations among three balanced random variables $X$, $Y$ and $Z$ that is relevant to this general setup \citep{3m2020} (p.\ 71):
\begin{align}
  \label{eq:genconst}
  \quad 1 - \rho_{XY}^2 - \rho_{XZ}^2 - \rho_{YZ}^2 + 2 \, \rho_{XY} \, \rho_{XZ} \,
  \rho_{YZ} \ge 0,\quad
\end{align}
where $\rho_{XY} := \frac{\langle XY \rangle}{\sigma_X\sigma_Y}$ is the \emph{Pearson correlation coefficient} for two balanced random variables $X$ and $Y$, $\langle XY \rangle$ is the covariance of $X$ and $Y$, and $\sigma_X$, $\sigma_Y$ are the standard deviations of $X$ and $Y$. Geometrically, it describes an inflated tetrahedron or \emph{elliptope} like the one depicted in Figure \ref{fig:elliptope}.\\

\section{Quantum Mechanics as a Natural Generalisation of Ordinary Causal Description}
\label{sec:natural_generalisation}

\subsection{The New Kinematics of Quantum Mechanics}
\label{sec:new_kinematics}

\begin{figure}[h]
   \includegraphics[width=2.5in]{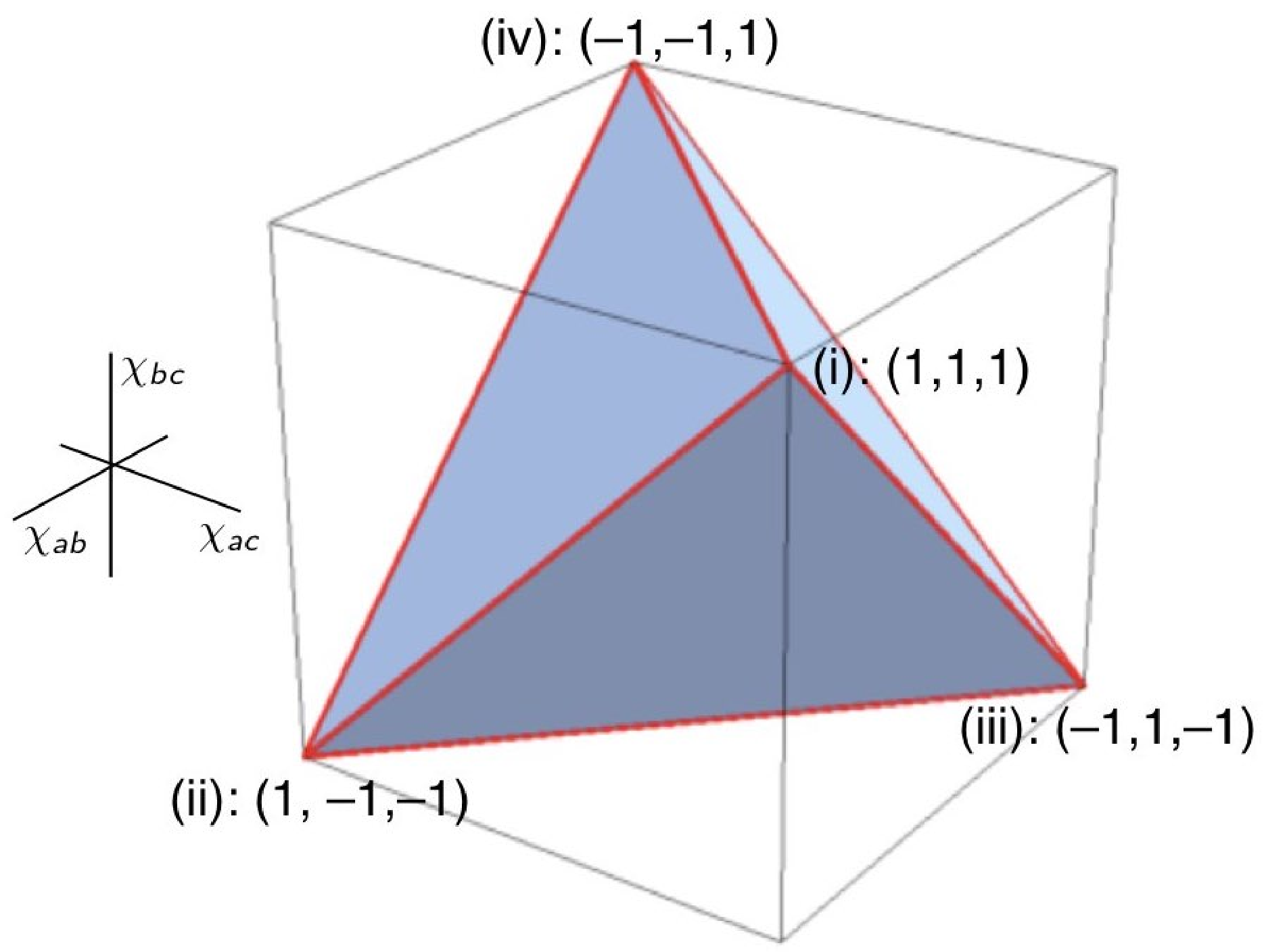} $\qquad\qquad$ \includegraphics[width=2.0in]{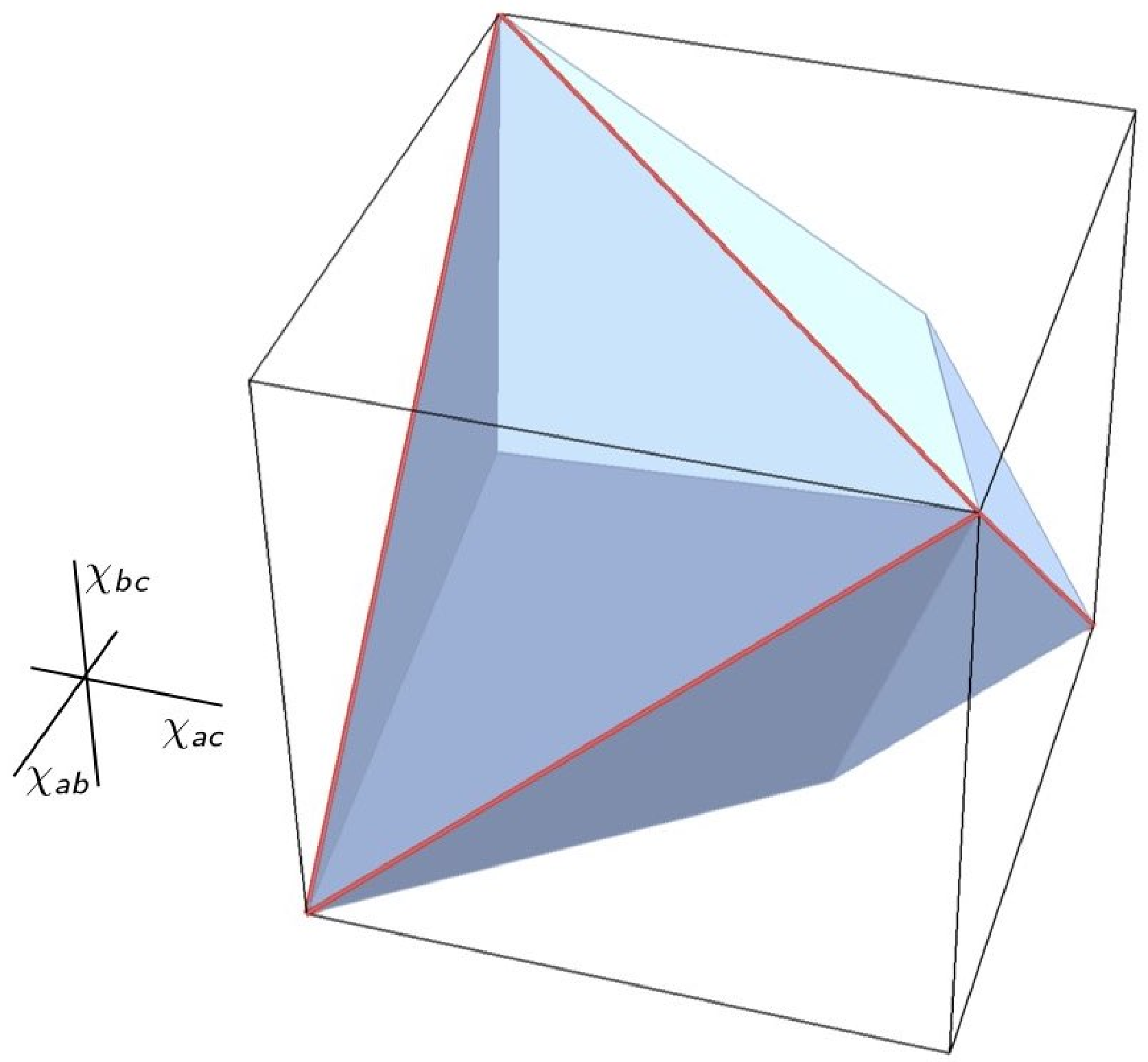} 
   \caption{On the (\textbf{left}): the tetrahedron of triplets of anti-correlation coefficients $(\chi_{ab}, \chi_{ac}, \chi_{bc})$ allowed by local hidden-variable theories in the Mermin-inspired setup (see the end of Section \ref{sec:necessary_conditions} above) for two possible values per balanced random variable. On the (\textbf{right}): the polyhedron corresponding to the case of three possible values. Note that whenever there are more than two values per variable, the corresponding correlation polytope will have more than three dimensions \citep{3m2020} (p.\ 106). The three-dimensional polyhedron at the right is a projection of this higher-dimensional polytope to three-dimensional space \citep{3m2020} (sct.\ 4.2.2). Image source: \citep{3m2020} (pp.\ 40, 140).\\}
   \label{fig:tetrahedron}
\end{figure}

The constraint given by Equation\ \eqref{eq:genconst}---which in \citep{3m2020} is called the \emph{elliptope inequality}---was known and discussed, albeit in contexts far removed from physics, as early as the 19th century by figures such as Udny Yule, Ronald A. Fisher and Bruno de Finetti \citep{3m2020} (chp.\ 3). As  Janas, Cuffaro \& Janssen explain, its derivation relies on the following fact about linear combinations of $X$, $Y$ and $Z$:
\begin{align}
  \label{eq:assump}
  \Big\langle \Big( v_1 \frac{X}{\sigma_X} + v_2 \frac{Y}{\sigma_Y}
    + v_3 \frac{Z}{\sigma_Z} \Big)^{\!2} \Big\rangle \ge 0,
\end{align}
where $v_1$ , $v_2$  and $v_3$ are real numbers. Modelling Equation\ \eqref{eq:assump} in a local hidden-variable theory requires a joint probability distribution over the possible values of $X$, $Y$ and $Z$. When there are two possible values per variable, the possible probabilistic correlations between $X$, $Y$ and $Z$ are describable geometrically as a tetrahedron (see Figure \ref{fig:tetrahedron}) lying entirely within the elliptope. When there are three or more values per variable, the associated polyhedra become further and further faceted and more closely approximate the elliptope, but become exceedingly difficult to compute. In contrast to a local hidden-variable theory, quantum theory (as John von Neumann observed in \citep{vonNeumann1927b}) allows us to assign a value to a sum of observable quantities---represented, for instance, by an operator $\hat{S} \equiv \hat{S}_X + \hat{S}_Y + \hat{S}_Z$---without, in general, requiring that we assign values to the individual summands $\hat{S}_X$, $\hat{S}_Y$ and $\hat{S}_Z$ (cf. \citep{stein1972} (p.\ 376)).  Janas, Cuffaro \& Janssen show how, as a consequence, the probabilistic correlations describable in quantum mechanics saturate the entire elliptope---already for \marginnote{\textbf{6}} spin-$\frac{1}{2}$ systems (the analogue of a two-valued variable) as well as for all higher values \mbox{of spin.}

That an assignment of values to a sum of observable quantities entails an assignment of values to the individual observables involved in the sum is always true in classical theory. By contrast, the kinematical structure of quantum theory---the constraints it imposes on our physical description of a system independently of the specifics of its dynamics (\citep{3m2020} \mbox{(chp.\ 1)}; \citep{janssen2009} (pp.\ 26--52))---is more general. Slogans such as ``quantum mechanics is all about information'' or ``quantum mechanics is all about probabilities'' are meant (at least for the informational approach under discussion here), not as ontological claims, but to emphasise that these constraints (as we will see in more detail in a moment) are probabilistic in nature---that the conceptual \emph{novelty} of quantum mechanics lies (\citep{3m2020} (sct.\ 6.3); \citep{demopoulosOnTheories} (sct.\ 4.3)) in the way that it constrains probability assignments. The slogan also conveys the idea that quantum mechanics is a framework that can in principle be applied to any type of physical system, for instance computational systems, the fictitious ``quantum bananas'' of \citep{bub2016} and so on (\citep{aaronson2013}; \citep{3m2020} (chps.\ 1, 6); \citep{nielsenChuang2000,wallace2019}).\\

Understanding why quantum mechanics, but not classical mechanics, saturates the elliptope for all values of spin is only one example of a problem that can be fully explained by appealing exclusively to quantum mechanics' kinematical constraints. In addition,  Janas, Cuffaro \& Janssen  \citep{3m2020} (sct.\ 6.4) describe three further examples of physical problems from the history of quantum mechanics that seemed, at first, to call for dynamical solutions but that were solved as a direct result of considering the change introduced into physics via quantum mechanics' novel kinematics. These are: accounting for the particle term in Einstein's 1909 formula for energy fluctuations in black-body radiation (pp.\ 188--189), accounting for the formula for the electric susceptibility of diatomic gases (pp.\ 189--191), and accounting for why electron orbits seem to depend on which coordinates one chooses to impose the quantisation condition (pp.\ 191--194).

The `measurement problem', as it is usually framed (see, e.g., \citep{myrvold2022}, p.\ 63), also seems to call for a dynamical solution, i.e., on the one hand, quantum mechanics describes the dynamics of systems as unitary in the absence of a measurement; yet, on the other hand, given a measurement the apparent `collapse' of the vector representing a system's state is non-unitary. But positing special measurement dynamics is ad hoc, or so the argument goes, since it should be possible to describe a measurement interaction as a physical interaction like any other. In fact, quantum mechanics agrees with this in the following sense: a quantum description of a measurement interaction can be provided, in unitary terms, to \marginnote{\textbf{7}} any desired level of detail given an appropriate placement of the `Heisenberg cut', on one side of which lies our quantum description of the interaction, and on the other, our classical description of the observation of its results. Stated in these terms, the `problem' can be seen as one of closing this gap, either by proposing a new theory to explain the connection between the two sides of the Heisenberg cut, or by arguing that quantum mechanics, in some sense, is all that one needs to make theoretical sense of it.

The Everett interpretation, as is well known, chooses the second option \citep{vaidman2018}. The same is true of the approach under discussion in this paper, though for very different reasons as we will soon see. The appearance of one rather than another of the possible outcomes of a measurement interaction is not thought of, in fact, in terms of a dynamical process at all---i.e., it is simply not taken to be described, in this sense, by quantum mechanics---let alone as a non-unitary dynamical collapse of the state vector. As Bub explains:

\begin{quote}
A unitary dynamical analysis of a measurement process goes as far as you would like it to go, to whatever level of precision is convenient. The collapse, as a conditionalization of the quantum state, is something you put in by hand after observing the actual outcome. The physics doesn't give it to you \citep{bub2016} (p.\ 228).
\end{quote}

There is, of course, more to be said. In particular, by considering how the kinematical structure of quantum mechanics departs from that of classical mechanics, one can discern two conceptually separate aspects of the so-called problem. In classical mechanics \citep{3m2020} (sct.\ 6.3), an observable $A$ is represented by a function, $f_A(\omega)$, defined on the phase space of a system. With $f_A(\omega)$, one can associate a Boolean algebra $\mathfrak{A}$, in which the possible yes-or-no questions concerning $A$ that can be asked regarding the system, questions of the form ``Is the value of the observable $A$ within the range $\Delta$?'' may be expressed. In classical mechanics, merely specifying a system's dynamical state, $\omega$, is enough to yield a determinate answer to every such question for every observable quantity associated with the system. In logical terms, this means that in classical mechanics, the Boolean algebras corresponding to each of the system's observables can be embedded within a globally Boolean algebra, such that a particular state assignment (which may or may not be probabilistic as, for instance, in classical statistical mechanics) suffices to fix the answers (or the probabilities over possible answers in the case of a probabilistic state assignment) to all of the possible questions that one can ask concerning any observable associated with the system. This is the sense in which the classical state is what \citet[]{bub-pitowsky2010} (p.\ 433) call a `truthmaker' in relation to a system's observables.\footnote{Note that the term `truthmaker' is intended here only in the logical sense described above, rather than in any metaphysical sense. For more on the use of the term `truthmaker' in philosophical contexts, see \citep{macbrideTruthmakerSEP}.}

In quantum mechanics \citep{3m2020} (sct.\ 6.3), an observable, $A$, is represented by a Hermitian operator, $\hat{A}$, acting on the Hilbert space associated with a system, whose possible values are given by the eigenvalues of $\hat{A}$. As with $f_A$ in the classical case, with $\hat{A}$ one can associate a Boolean algebra $\mathfrak{A}$ representing the possible yes-or-no questions that one can ask about $A$. But the quantum state, unlike the classical state, fails to be a `truthmaker' in relation to a system's observables in two ways, corresponding to what I will be calling the `big' and the `small' measurement problems. The `big problem' is that unlike in classical mechanics, where one can always in principle eliminate the indeterminacy associated with any given probabilistic phenomenon by including further parameters in one's dynamical model of the system of interest, in quantum mechanics, fully specifying the normalised vector $| \psi \rangle$ representing the state of the system of interest can only ever yield the probability, which is in general neither 0 nor 1, that the answer to a given experimental question will take on a particular value. Given that it will nevertheless be possible, conditional on the selection of an observable to measure, to describe the observed relative frequencies of the various possible outcomes of the (by assumption, projective) measurement in terms of a classical probability distribution as given by the Born rule, this departure from classicality is arguably only minor, at least in comparison with the `small measurement problem', the more significant way in which the kinematics of quantum mechanics diverges from classical mechanics. This refers to the fact that the classical probability distributions that can be \marginnote{\textbf{8}} associated with the system's observables, in the way just described, cannot be embedded into a global classical probability distribution over all of the system's observables, or alternately that the Boolean algebras corresponding to each of the system's observables cannot be embedded within a globally Boolean algebra. Moreover, quantum mechanics' unitary description of a measurement interaction does not, by itself, prefer any one of these classical (i.e., Boolean) ways of effectively characterising the system \citep{3m2020} (p.\ 224). In the next section, we will consider what to say about the significance of this on the informational, or (neo-)Bohrian, approach under discussion in this paper.\footnote{Note that the distinction between a `big' and a `small' measurement problem was first introduced by \citet[]{pitowsky2006}, and is further developed in \citep{bub-pitowsky2010}, \citep{bub2016} (in the first edition) and in \citep{3m2020}. \citet[]{brukner2017} also distinguishes between the same two aspects of the measurement problem but inverts the labels and does not use inverted commas. Here, I am following \citep{3m2020}. Note that although the `big' and especially the `small' problems are formulated somewhat differently in \citep{3m2020} than in \citep{bub-pitowsky2010}, I take there to be no difference of substance. Although Bub and Pitowsky's idea that the `small' problem is resolved by ``considering the dynamics of the measurement process and the role of decoherence in the emergence of an effectively classical probability space of macroevents to which the Born probabilities refer'' \citep{bub-pitowsky2010} (p.\ 438) is not incorrect, it is (in my view) misleading insofar as the formulation seems to suggest that this, by itself, is enough to yield an observer-independent description of the dynamics of a measurement interaction. Whether a dynamical analysis to this end can be made to work is an interesting question (for further discussion, \mbox{see \citep{bacciagaluppi2020, crull2022}}). But that this question is not Bub and Pitowsky’s  is evident, in any case, later in the same article where they explain that the goal of such a dynamical analysis is to provide ``a \emph{consistency proof} that the familiar objects of our macroworld behave dynamically in accordance with the kinematic probabilistic constraints on correlations between events.'' \citep{bub-pitowsky2010} (p.\ 452, emphasis mine). In other words, it is taken for granted, in any such analysis, that specific observables have been selected for measurement that have particular phenomena associated with them, or as Bub has put it in his more recent publications, it is required that one posits some ``ultimate measuring instrument'' that is not included in one's quantum-mechanical description of a measurement interaction \citep{bub2018a} (pp.\ 8--9). In my terms this amounts to a Boolean frame. Note that the `small measurement problem' is what Everettians call the `preferred basis problem'. It is also what is highlighted by the (in)famous thought experiment of \citet{frauchigerRenner2018}. For some recent commentaries on this and similar thought experiments that are directly relevant to the approach to quantum mechanics under discussion in this paper, see \citep{bub2021, dascal2020, felline2020a}.}

\subsection{The Subjective Character of the Idea of Observation--Schematising the Observer as a Postulate}
\label{sec:subjective_character}

On what I will call the \emph{traditional metaphysical picture}---the one lying behind Bell's insistence that in any physical theory worth its salt, ``[o]bservables are \emph{made} out of beables'' \citep{bell1973} (p.\ 41, emphasis in original)---the possible values of dynamical variables like position, momentum, the direction of a particle's spin and so on, are understood to be the manifestations of an underlying reality whose properties are revealed in our experiments with physical systems.\footnote{Note that, in the sequel, I will for the most part be using the term property rather than beable, in part to convey that the traditional metaphysical picture I am referring to here transcends quantum mechanics.} The trouble with quantum mechanics, given this picture, is that because of the big and especially the small `measurement problems', the possible values of dynamical quantities---represented, in quantum mechanics, as the eigenvalues of a given Hermitian operator (cf. \citep{acuna2021, bub2010b, dieks2017})---cannot be taken to represent determinate properties of a single classically describable physical system in the following logical sense: there is no globally Boolean algebra in which one can embed all of the individual Boolean algebras corresponding to the various observables associated with the system, that can be used to derive, given a state assignment, an unconditional probability distribution over the possible values of a given observable, let alone a determinate value. Given the traditional metaphysical picture, there are, broadly speaking, two attitudes that one can take towards quantum mechanics. First, one can take it to be incomplete and pursue a \marginnote{\textbf{9}} research programme to complete it by positing further, perhaps unobservable, physical parameters not described by quantum mechanics that can be used to provide us with an absolute representation of a system in some sense (see, \mbox{e.g., \citep{ghirardi2018, goldstein2006}}). Alternately, one can insist that quantum mechanics already provides us with a globally Boolean picture of the world---but that that world is a multiverse rather than the single classically describable universe we at one time imagined it to be \citep{vaidman2018}.

As for the approach under discussion here (cf. Stein's \citep{stein1972} (pp.\ 369, 409--410) distinction between what he calls the epistemological and metaphysical senses of interpretation), it is not opposed to the traditional metaphysical picture per se. That picture would be apt if classical mechanics (or some other theoretical framework with a globally Boolean algebra of observables) were our fundamental framework. But it is also open to the possibility that that picture is not apt; for how one carves nature at the joints, so to speak, is something that, for the informational interpreter, should be motivated by physical theory rather than a priori. The informational interpreter is certainly committed to something a priori, but it is not a particular metaphysical thesis about the way the world must be. What the informational, or (neo-)Bohrian, interpreter is committed to, rather, is the \emph{empiricist methodology} through which one reasons, from the values revealed in experiments with what we take to be physical systems, carried out under precisely specified---to the relevant scale and for the relevant purposes (cf. \citep{curiel2020} (sct.\ 5, sct.\ 11, point 5))---experimental conditions, to a picture of the world that is anchored in the dynamical model one builds of the phenomena in a given context.\footnote{Note that the local hidden-variable models (in the form of raffles), for spin correlations described in \citep{3m2020} (chps. 2--3), are (toy) examples of such models. To say that an overarching theoretical picture must be anchored in a toy model like this is to say that that toy model must remain valid when one restricts the phenomena under consideration to those associated with the experimental context that the toy model was designed to characterise (cf. \citep{perovic2021} (pt. 2)).}

The approach under discussion is not a `realist' one in the sense that ultimately it is taken to be the goal of a physical theory---even a so-called fundamental physical theory---to represent \emph{phenomena} rather than the so-called reality that one might imagine to be in some sense responsible for the phenomena, in a systematic way (\citep{bub2016} (p.\ 227); \citep{demopoulosOnTheories} (pp.\ 135--139); \citep{3m2020} (pp.\ 219--222); \citep{pitowsky1994} (pp.\ 111, 118); cf. \citep{stein1989} (p. 50); \citep{stein1994} (pp.\ 639, 645)).\footnote{Note that, in this context, the word `fundamental', of course, cannot be construed as pertaining to `fundamental stuff'; rather, a (candidate) fundamental theory should be understood, for one defending the approach that we are discussing here, as one that we take to be capable of representing all known phenomena.} Insofar as one does this for the purposes of using physical theory as a tool, the approach can be called instrumentalist (cf. \citep{adlam2022} (p.\ 2)). But instrumentalism in this sense is compatible with realism on a more reasonable (methodological) construal of what the latter means. To put it succinctly, the important question on this approach to quantum mechanics is not whether but \emph{how} to use physical theory to assign physical properties to what one takes to be the system of interest responsible for a given phenomenon (\citep{3m2020} (\mbox{pp.\ 8--10} and chp.\ 6); \citep{perovic2021} (p.\ 118); cf. \citep{stein1972} (p.\ 371)).

One could perhaps conceive of a more radical form of empiricism that is not even realist in this methodological sense. Although this is, presumably, a logically possible attitude to take to science, I confess that I struggle to imagine what a science that followed such a methodology would be like. For the methodological realism advocated on the approach we are discussing here amounts to the demand \citep{bohr1958} (p.\ 310) that we be able to meaningfully account to one another how we have set up a particular experiment (``what we have done''), and what information it yields (``what we have learned'') about an object that we model as able to interact with our experimental apparatus in a particular way \citep{perovic2021} (pp.\ 44--45). This is the methodology characteristic of what Bohr called the ``ordinary causal description'' of phenomena that a framework like classical mechanics makes precise and for which quantum mechanics provides a generalisation. And it is in this methodological sense that the ordinary causal description functions as a fundamental constraint on the approach to quantum mechanics under discussion.

\marginnote{\textbf{10}} Stein famously suggested that the principal difficulty in making sense of the connection between the `observational' and `theoretical' parts of a physical theory is that of how to account, theoretically, for observation, or as he puts it, ``how to get the laboratory inside the theory.'' \citep{stein1994} (p.\ 638). This issue, for Stein, is of the highest importance, for ``[i]t would $\dots$ be impossible to \emph{understand} a theory, as anything but a purely mathematical structure---impossible, that is, to understand a theory \emph{as} a theory of physics---if we had no systematic way to put the theory into connection with observation (or experience).'' \citep{stein1994} (p.\ 639).

Stein observes that Carnap's approach to the question, which assumes that the connection between the theoretical and observational parts of language (at least in physics) is deductive, faces a fundamental barrier insofar as (according to Stein) that assumption is de facto false: ``there is no department of fundamental physics in which it is possible, in the strict sense, to \emph{deduce} observations, or observable facts, from data and theory.'' \citep{stein1994} (p.\ 638). Instead, Stein suggests that the way that theory and experiment are connected is by ``schematizing the observer within the theory'' (\citep{stein1994} (p.\ 649); cf. \citep{stein1972} (sct.\ XVII)). Curiel elaborates on the idea:

\begin{quote}
  We need a way to understand the substantive, physically significant contact---the epistemic continuity, as it were---between a precisely characterizable situation in the world of experience and the mathematical structures of what we usually think of as our theories. Such understanding should at a minimum consist of an articulation of the junctions where meaningful connections can be made between the two, and would thus ground the possibility of the epistemic warrant we think we construct for our theories from such contact and connection. \citep{curiel2020} (p.\ 6).
\end{quote}

``By `schematize the observer','' Curiel writes, ``I mean something like: in a model of an experiment, to provide a representation of something like a measuring apparatus, even if only of the simplest and most abstract form, that allows us to interpret the model \emph{as} a model of an experiment or observation.'' \citep{curiel2020} (p.\ 9). As one of a number of physical examples he uses to motivate the idea, Curiel considers the case of the International Practical Temperature Scale, originally specified in \citep{burgess1928} and revised in subsequent decades, the goal of which is to represent the ideal thermodynamic temperature scale as closely as possible throughout its range. As Curiel explains, between the primary fixed point 0.01$^{\mathrm{\textdegree}}$C and the secondary fixed point 630.5$^{\mathrm{\textdegree}}$C, the scale may be defined in terms of the Callendar equation which relates temperature to the resistance of platinum as measured when immersed in various media and taking into account the particular features of the kind of platinum being used. Importantly, the variables related by the Callendar equation are defined, in part, by interpolation on the basis of our knowledge of measurements carried out at a particular scale \citep{curiel2020} (p.\ 13). In particular, below and above these two fixed points, the Callendar equation diverges from the thermodynamic scale, requiring other measurable quantities to be taken into account. Curiel takes this to show that ``one cannot even define physical quantities---e.g., temperature---without explicit schematic representation of the observer, much less have understanding of how to employ their representations in scientific reasoning in ways that respect the regime of applicability.'' \citep{curiel2020} (p.\ 14).\footnote{Note that it does not follow from the fact that one requires of a physical concept that it have operational significance that one is committed to operationalism in the strong sense sometimes attributed to Percy Williams Bridgeman \citep{chang2021}. Requiring that a concept have operational significance to \emph{qualify} as a physical concept is not the same thing as taking a given set of operations to exhaust its meaning, or requiring that a given concept be uniquely specified by a given set of operations (see also \citep{myrvold2010}, (sct.\ 9.3); \citep{myrvold2011}; \citep{myrvold2021}, (pp.\ 141--145)). The density operator, for instance, clearly does not satisfy the latter criterion insofar as there are infinitely many given state preparations that correspond to a given density operator whenever the latter is not pure. I will have more to say about the physical significance of the density operator in the next section.}

Curiel's general point was one that was well understood by Bohr. In his aforementioned `Como paper', commenting on the use of the superposition principle to explain \marginnote{\textbf{11}} particle-like quantum phenomena in terms of the concept of a `wave packet', Bohr pointed out that:

\begin{quote}
Rigorously speaking, a limited wave-field can only be obtained by the superposition of a manifold of elementary waves corresponding to all the values of $\nu$ and $\sigma_x$, $\sigma_y$, $\sigma_z$. But the order of magnitude of the mean difference between these values for two elementary waves in the group is given in the most favourable case by the condition

\begin{equation}
  \label{eq:bohr1}
\Delta t \Delta \nu = \Delta x \Delta \sigma_x = \Delta y \Delta \sigma_y = \Delta z \Delta \sigma_z = 1
\end{equation}
 where $\Delta t, \Delta x, \Delta y, \Delta z$ denote the extension of the wave-field in time and in the direction of space corresponding to the co-ordinate axes \citep{bohr1928} (p.\ 581).
\end{quote}

Here, $\nu$ refers to the frequency, and $\sigma_x, \sigma_y, \sigma_z$ refer to the wavenumbers for the elementary waves in the directions of the coordinate axes. All else equal, the broader the spread of wavenumbers/frequency in the wave group, the more determinate the spatiotemporal extent of the resultant packet, and vice versa. Now, according to the de Broglie relations, $E = \hbar\nu$, $I = \hbar\sigma$, where $\hbar = h / 2\pi$ is the reduced Planck's constant. If we multiply Equation \eqref{eq:bohr1} by $\hbar$, this gives us Heisenberg's uncertainty relations: 
\begin{equation}
\Delta t \Delta E = \Delta x \Delta I_x = \Delta y \Delta I_y = \Delta z \Delta I_z = \hbar
\end{equation}
which give the upper bound on the accuracy of momentum/position determinations with respect to the wave-field.

As the wave-field associated with the object gets smaller---thus allowing us to `zoom in', so to speak, on its position and time coordinates---the possibility of precisely defining changes in the energy and momentum associated with the object decreases in proportion. And the opposite is also true: given a larger wave-field, it will be possible to `zoom out' for the purposes of a determination of the object's momentum (or energy), but in this case, one foregoes a precise definition in relation to the object's spatiotemporal coordinates. Note that `zooming in' and `zooming out' are associated with different experimental arrangements. For the case of a $\gamma$-ray microscope, they are associated with the finite size of the microscope's aperture. The indeterminacy in our assignments of position and momentum to the system is not due to the interaction between the object and the measuring apparatus per se, but to the fact that certain experimental arrangements, well-suited for precisely determining momentum, make it such that in the limit, it becomes impossible \emph{to define} changes in the object's spatiotemporal coordinates, and vice versa. Bohr sums this up as follows:

\begin{quote}
Indeed, a discontinuous change of energy and momentum during observation could not prevent us from ascribing accurate values to the space-time co-ordinates, as well as to the momentum-energy components before and after the process. The reciprocal uncertainty which always affects the values of these quantities is, as will be clear from the preceding analysis, essentially an outcome \emph{of the limited accuracy with which changes in energy and momentum can be defined}, when the wave-fields used for the determination of the space-time co-ordinates of the particle are sufficiently small \citep{bohr1928} (p.\ 583. emphasis mine).
\end{quote}

Quantum mechanics, on the approach we are discussing, is to be understood as elevating the insight, which Stein and Curiel (as we discussed above) have referred to as the practical and epistemic necessity---for understanding a theory as a theory of physics---of ``schematising the observer,'' to the level of a postulate (cf. \citep{hansenWolfFeatureNotBug}). Bohr was explicit \mbox{about this:}

\begin{quote}
In the treatment of atomic problems, actual calculations are most conveniently carried out with the help of a Schr\"odinger state function, from which the statistical laws governing observations obtainable under specified conditions can be deduced by definite mathematical operations. It must be recognized, however, that we are here dealing with a purely symbolic procedure, \emph{the unambiguous physical} \marginnote{\textbf{12}} \emph{interpretation of which in the last resort requires a reference to a complete experimental arrangement.} Disregard of this point has sometimes led to confusion, and in particular the use of phrases like `disturbance of phenomena by observation' or `creation of physical attributes of objects by measurements' is hardly compatible with common language and practical definition. \citep{bohr1958} (pp.\ 392--393, my emphasis).
\end{quote}

On the (neo-)Bohrian approach under discussion here, an observer is represented by a `Boolean frame' \citep{3m2020} (p.\ 213)---the Boolean algebra within which one represents the possible yes-or-no questions concerning a given observable, $A$, that can be asked about the system of interest: questions of the form ``Is the value of $A$ within the range $\Delta$?''. Given the schematic representation, to the relevant scale and for the relevant purposes, of an observer in this sense, one may then use the language of quantum mechanics to give a physical analysis, in terms of the states of two interacting dynamical systems, $\mathcal{S}$ and $\mathcal{M}$ (representing the measuring device), of how the observed relative frequencies of outcomes of assessments of $\mathcal{M}$ will be (assuming the measurement is ideal) describable using a particular classical probability distribution over possible values of $A$ that can be thought of as determined in conformity with the dynamics of $\mathcal{S}$ and $\mathcal{M}$ \citep{3m2020} (pp.\ 202--212). In classical mechanics, because the state is a truthmaker in the sense we discussed in Section \ref{sec:new_kinematics}, \emph{as a matter of logic} one can always argue that any given schematisation of the observer in the above sense is superfluous, at least in principle.\footnote{Note that what I mean by superfluous is that the question: ``Is the value of $A$ within the range $\Delta$?'' is equivalent, in classical mechanics, to the question of whether $f_A(\omega) \in \Delta$ \citep{hughes1989} (p.\ 61), where $f_A(\omega)$ is the function defined on the system's phase representing the observable $A$ (see Section \ref{sec:new_kinematics}).} This is not the case in quantum mechanics, where the imposition of a Boolean frame is required in order to interpret the outcome of a measurement interaction as providing us with information about the world.

\subsection{\sloppy The Classical Idea of Isolated Objects and the Quantum-Mechanical Concept of an \mbox{Open System}}
\label{sec:isolated_objects}

If it is only ever possible to describe one's experience \emph{as} the experience of a system in the context of an interaction between what one takes to be that system and something else, then the system that one takes oneself to be describing is in every case an open system.\footnote{For more on the physics and philosophy of open quantum systems, see \citep{cuffaroHartmannOpenSystemsView}.} In an article published in \emph{Dialectica}, in which he argued that quantum mechanics should be judged to be incomplete, Albert Einstein wrote:

\begin{quote}
Without $\dots$ an assumption of the mutually independent existence (the 'being-thus') of spatially distant things, an assumption which originates in everyday thought, physical thought in the sense familiar to us would not be possible. (\citep{einstein1948}, as translated by \citet[]{howard1985} (p.\ 187)).
\end{quote}

This passage, and the wider argument of which it is a part, has been much commented on (see, for instance, \citep{demopoulosOnTheories} (chp.\ 4); \citep{howard1985,ramirez2020}). Here, I only want to point out that it amounts to the demand that we be able to treat spatially separated subsystems of the universe as isolated (cf. \citep{wallaceIsolatedSystemsI}), and that arguably, we should construe this as a methodological demand---a claim about what we must be able to assume if we are to be able to practice physics in the sense familiar to us at all---rather than an a priori claim about how the world is.\footnote{In any case it can be plausibly read in this way given Einstein's career-long concern with methodological issues (see \citep{lehner2014}). For a discussion of Einstein's mutually independent existence condition, in particular, that interprets it in this way, see \citep{demopoulosOnTheories} (chp.\ 4) and also \citep{cuffaroInfCausality}.} As with Bohr, my goal here is not Einstein exegesis. Irrespective of what he actually understood himself to be saying in this passage, the idea of understanding the ``assumption of mutually independent existence'' of spatially distant things as a methodological requirement on physical inquiry is prima facie plausible. And understanding how the principle, construed in this way, is (\emph{pace} Einstein) satisfied by quantum mechanics, illuminates important aspects of the informational approach under discussion in this paper.

\marginnote{\textbf{13}} Besides allowing us to express that a given system has been prepared in one of a number of states, $\{| \psi_i \rangle\}$, with probabilities $\{p_i\}$, a density operator like
\begin{align}
  \label{eqn:density_operator}
  \rho = \sum_i p_i| \psi_i \rangle\langle \psi_i |,
\end{align}
where $| \psi_i \rangle\langle \psi_i |$ is the projection operator associated with the state vector $| \psi_i \rangle$, is also used in quantum mechanics to represent the state of an open system, by which I mean one dynamically evolving under the influence of an external `environment': for instance, a measurement device $\mathcal{M}$ that has interacted with the system. When it refers to an open system, a density operator like the one given in Equation\ \eqref{eqn:density_operator} is said to represent an `improper' mixture \citep{despagnat1966, despagnat1976} of the `pure states', $\{| \psi_i \rangle\}$---improper because, owing to the fact that $\mathcal{S}$ and $\mathcal{M}$ are entangled, it is actually impossible to interpret Equation\ \eqref{eqn:density_operator} as literally describing a system that is in a given pure state $| \psi_i \rangle$ with a given probability $p_i$ (because a subsystem of an entangled system can never be in a pure state).

Of course, if $\mathcal{S}$ and $\mathcal{M}$ were not entangled, we could interpret Equation\ \eqref{eqn:density_operator} as representing our ignorance regarding the actual state, $| \psi_i \rangle$, that the system is in. In this case, we would say that the density operator represents a `proper' mixture of the pure states $\{| \psi_i \rangle\}$ (although even in this case such statements should be taken with a grain of salt, because for a given ensemble whose state is represented by some density operator $\rho$, there are in general infinitely many preparation procedures that will give rise to it, i.e.,
$\rho = \sum_j p_j \, | \psi_j \rangle \langle \psi_j | = \sum_k p'_k \, | \phi_k \rangle \langle \phi_k |$, whenever $\sum_j p_j \, | \psi_j \rangle \langle \psi_j |$ and $\sum_k p'_k \, | \phi_k \rangle \langle \phi_k |$ are related by a unitary transformation \citep{nielsenChuang2000} (p.\ 103)). These two physical situations (i.e., those represented by an improper and a proper mixture, respectively) are not the same. Nevertheless, a sequence of measurements on the members of an improperly mixed ensemble will be effectively indistinguishable from---in the sense that they will be described by the same probability distribution as---a sequence of those same measurements on a properly mixed ensemble whose state is also described by $\rho$. In the context of a consideration of spatially separated systems, this amounts to the `no-signalling' condition (a misnomer as it is not a relativistic constraint per se), which asserts that the marginal probabilities associated with outcomes of local experiments on a subsystem of any quantum-mechanically described system are independent of whatever particular experiments are performed (or whether any are performed at all) on the other subsystems. This effectively means that we can treat physical systems in different regions of space \emph{as if} they had mutually independent existences for the purposes of experiments local to those regions (\citep{cuffaroInfCausality}; \citep{demopoulosOnTheories} (chp.\ 4); cf. \citep{wallace2010a}). It is important to emphasise that the existence of nonlocal correlations is not being denied. Instead, what is being affirmed is the fact that according to quantum mechanics, we can learn about them using local means.

On the informational approach we are discussing, a quantum state description is \emph{not} taken to represent a property or a collection of properties that one can think of as possessed by a system independently of a given experimental context, for it is precisely the experimental context, represented as a Boolean frame, that allows one to give an account, consistently with a given state assignment, of how the experimental apparatus involved dynamically interacts with a system, thus allowing us to conceive of some phenomenon as representing a value of a given property of the system in the first place. What the quantum state \emph{does} represent is the structure of and interdependencies among the possible ways in which one can give a probabilistic characterisation of a system in the context of a physical interaction (\citep{3m2020} (p. 186); \citep{cuffaroHartmannOpenSystemsView} (pp.\ 19--20)). A classical state is no different in this sense.\footnote{Consider, for instance, Curiel's construal \citep{curiel2014} (sct.\ 3) of the configuration space of an abstract classical system as encoding a description of its kinematically possible interactions with other abstract classical systems.} But because the probability distributions over the values of every classical observable are determined by the state independently of whether a physical interaction through which one can assess those values is actually made, there is an invitation to think of them as originating in the properties of an underlying physical system that exists in a particular way irrespective of anything external.\footnote{At least this is true when one considers the situation abstractly, and in particular, when one disregards arguments along the lines of Curiel's and Stein's that should make one skeptical about whether it actually makes sense, even in classical physics, to speak in such absolute terms.} \marginnote{\textbf{14}} Although it is not denied that one can make such a picture work (along the lines we discussed at the beginning of Section \ref{sec:subjective_character}) if one really wants to \citep{3m2020} (pp.\ 229--230), the more complex structure of observables related by quantum mechanics does not similarly invite the inference from the values of observable quantities to the properties of an underlying system in the sense that there is no globally Boolean frame that one can use to characterise all of a given system's observables. And since the informational interpreter is not committed to seeking a globally Boolean picture, she is not committed to the project of making such a picture work \emph{in spite} of quantum mechanics.

The elliptope and polyhedra depicted in Figures \ref{fig:elliptope} and \ref{fig:tetrahedron} are a way to visualise, in the general setup I introduced in Section \ref{sec:necessary_conditions}, the sense in which local hidden-variable theories are able to represent only a special case of the phenomena that more general frameworks like quantum mechanics can represent. But just as in classical mechanics, in a given measurement context that we can---by assumption---effectively describe in Boolean terms, one can, consistently with quantum mechanics, provide a dynamical model of the measurement interaction also in such terms (e.g., some mixture of the classical raffles discussed in \citep{3m2020} (chps. 2--3)). Such a model will not `suffer from the small measurement problem' (since the observables associated with that measurement context commute). As for the `big measurement problem', the short answer is that for the informational interpreter, one simply accepts it as a brute fact that nature is indeterministic \citep{3m2020} (p.\ 11). But if one insists on a deterministic model, then the informational interpreter will point out that in any given measurement context (and associated Boolean frame), it will always be possible to interpret the indeterminacy of individual measurement results, in a given experimental run, as if they stem from our inability to completely specify some relevant physical parameter in the model.

But can nothing really be said, on this informational view, about what the world is like independently of observation? On the contrary, our assignments of values to non-dynamical quantities like mass, spin and charge are valid irrespective of the experimental context they are relevant to (\citep{demopoulosOnTheories} (p.\ 184); \citep{3m2020} (p.\ 217)). Regarding dynamical quantities, one may say that the world is such that all of the effectively classical (i.e., Boolean) pictures that one can draw of it, under the precisely specified experimental conditions corresponding to each of them, are precisely relatable to one another, probabilistically, in a way that is necessarily constrained by the kinematical structure of quantum mechanics. Neither of these statements is trivial. But one may nevertheless wonder (assuming one finds this to be objectionable), whether the second truth somehow depends upon the actual existence of conscious observers. The (neo-)Bohrian will answer no.\footnote{Note that this is, I think, one way to distinguish (in the sense of where their main focus lies) the view under discussion in this paper from some of the other approaches to interpretation that we mentioned at the beginning of Section \ref{s:introduction} like QBism \citep{fuchs2017}.} Rather, in describing the structure of our world in this way, a schematic representation of what relevantly constitutes an observer---a Boolean frame---is used as a formal tool with which to describe how the various dynamical possibilities, or `propensities' \citep{3m2020} (p.\ 218), that are implicit in the world necessarily relate to one another. A particular Boolean frame acquires physical significance through the specification, which can be given using the language of quantum mechanics, of a dynamical model of how the associated measuring apparatus interacts with a given system, but the specification of a given context in this sense in no way implies that it must actually be instantiated or actually be interpreted by anyone; it only specifies (schematically) \emph{how} to do so.\footnote{Note that the term `propensities' should not necessarily be taken in the sense of the interpretation of probability first formulated and defended by Karl Popper \citep{popper1959}, but only to signify that (as is also the case in Popper's interpretation) the probabilities for outcomes of experiments can be thought of as determined given a specification of the experimental setup. Since (as I explained in Section \ref{sec:subjective_character}) a subjective component is necessarily involved in one's characterisation of a given setup, however, and since, given an experimental setup, one may (as I mentioned in the previous paragraph) interpret the indeterminacy of individual measurement results in a given run as if they were due to our incomplete knowledge of some relevant physical parameter, it would arguably be more appropriate to think of the probabilities of outcomes of measurements in a given context as if they were more akin to what Wayne Myrvold \citep{myrvold2021} (pp.\ 106--121, 209--210) has called `epistemic chances' rather than to Popper's propensities.} \marginnote{{\normalsize \textbf{15}}}

\section{The View in a Nutshell}
\label{sec:nutshell}

On the informational, or (neo-)Bohrian, approach that concerns us here, quantum mechanics is about probabilities. These are understood to be (to use von Neumann's phrase) ``given from the start'' (quoted in \citep{bubForewordToRaffles} (p.\ x)), i.e., as objectively associated with a given precisely specified, to the relevant scale and for the relevant purposes, experimental context representable as a Boolean frame. Quantum mechanics describes the relations between these in an in general non-Boolean way, which amounts to saying that the various probability distributions that one can use to effectively characterise the phenomena associated with commuting sets of observables cannot be embedded into a global probability distribution over the simultaneous values of all observables. Despite this, quantum mechanics provides a recipe through which one can acquire information concerning a system through interactions with objects whose relevant parameters can---effectively---be described using classical, i.e., \emph{Boolean}, means, as being either ``on'' or ``off'' with a certain probability determined by the dynamical properties of the system according to the dynamical model that one constructs of it in that experimental context. In other words (\emph{pace} Einstein), \emph{quantum mechanics allows us to do physics in much the same way as we always have} (Bub, personal communication). But it does not follow from any of this---\emph{the `measurement problem' is a feature}, not a bug---that nature itself must be such as to allow (in a natural way, at any rate) for a globally Boolean description of all aspects of all dynamical phenomena that physics is concerned to describe \mbox{(cf. \citep{pitowsky1994} (p.\ 118))}.

\vspace{6pt} 

\section*{Funding}

This research was funded by the Alexander von Humboldt Foundation and by the German Research Council (DFG) through grant number 468374455.

\section*{Acknowledgements}

Thanks to Jeff Bub, Erik Curiel, Michael Janas, Michel Janssen, Slobodan Perovi\'c and two anonymous referees for their comments on a previous draft. This paper also benefited from discussion with Emily Adlam, Guido Bacciagaluppi, Philipp Berghofer, John De Brota, Robert DiSalle, Sam Fletcher, Chris Fuchs, Bill Harper, Stephan Hartmann, Philip Goyal, Richard Healey, Leah Henderson, G\'abor Hofer-Szab\'o, Philipp H\"ohn, Michael Kiessling, Ravi Kunjwal, Samo Kuto\v{s}, James Ladyman, Fred Muller, Wayne Myrvold, Alyssa Ney, Pascal Rodr\'iguez-Warnier, Simon Saunders, R\"udiger Schack, Paul Teller, Iulian Toader, Lev Vaidman and the members of the philosophy of physics reading group organised by John Dougherty at the Munich Center for Mathematical Philosophy in the summer 2022 term.


\begin{thebibliography}{999}
    
\bibitem[{Demopoulos(2022)}]{demopoulosOnTheories}
Demopoulos, W. 
\newblock {\em On Theories\/}; 
\newblock  Harvard University Press: Cambridge,  MA, USA, 2022.


\bibitem[{Perovi\'c(2021)}]{perovic2021}
Perovi\'c, S. 
\newblock {\em From Data to Quanta--Niels Bohr's Vision of Physics\/}; 
\newblock  University of Chicago Press: Chicago, IL, USA, 2021.


\bibitem[{Evans(2020)}]{evans2020}
Evans, P.W. 
\newblock Perspectival objectivity.
\newblock {\em Eur. J. Philos. Sci.} \textbf{2020}, {\em 10}, 19.
 

\bibitem[{Brukner(2017)}]{brukner2017}
Brukner, {\v{C}}. 
\newblock On the quantum measurement problem.
\newblock In {\em Quantum [Un] Speakables {II}\/}; Springer:  Berlin/Heidelberg, Germany, 2017; pp.\ 95--117.


\bibitem[{Bub(2017)}]{bub2017}
Bub, J. 
\newblock Why {B}ohr was (mostly) right. \emph{arXiv} \textbf{2017},
\newblock {arXiv:1711.01604}.


\bibitem[{Janas et~al.(2022)Janas, Cuffaro, \& Janssen}]{3m2020}
Janas, M.; Cuffaro, M.E.;  Janssen, M. 
\newblock {\em Understanding Quantum Raffles: Quantum Mechanics on an
  Informational Approach: Structure and Interpretation\/};
\newblock Springer: Cham, Switzerland,  
\newblock 2022.


\bibitem[{Landsman(2017)}]{landsman2017}
Landsman, K. 
\newblock {\em Foundations of Quantum Theory\/};
\newblock  Springer: Cham, Switzerland,   2017.


\bibitem[{Fuchs(2017)}]{fuchs2017}
Fuchs, C.A. 
\newblock Notwithstanding {B}ohr, the reasons for {QB}ism.
\newblock {\em Mind Matter} \textbf{2017}, {\em 15}, 245--300.


\bibitem[{Healey(2017)}]{healey2017}
Healey, R. 
\newblock {\em The Quantum Revolution in Philosophy\/}; 
\newblock Oxford University Press: Oxford, UK, 2017.


\bibitem[{Rovelli(2021)}]{rovelli2021}
Rovelli, C. 
\newblock {\em Helgoland: Making Sense of the Quantum Revolution\/}; 
\newblock  Riverhead Books: New York, NY, USA, 2021.


\bibitem[{Pitowsky(1989)}]{pitowsky1989}
Pitowsky, I. 
\newblock {\em Quantum Probability---Quantum Logic\/}; 
\newblock  Springer: Hemsbach, Germany, 1989.


\bibitem[{Bub \& Pitowsky(2010)}]{bub-pitowsky2010}
Bub, J.; Pitowsky, I. 
\newblock Two dogmas about quantum mechanics.
\newblock In  {\em Many
  Worlds? {E}verett, Quantum Theory, and Reality\/}; Saunders, S.,  Barrett, J., Kent, A.,  Wallace, D.,  Eds.;  
  Oxford University Press: Oxford, UK,  2010; pp.\ 433--459.


\bibitem[{Bub(2016)}]{bub2016}
Bub, J. 
\newblock {\em Bananaworld, Quantum Mechanics for Primates\/}, 2nd paperback ed.; 
\newblock Oxford University Press: Oxford, UK,  2016.


\bibitem[{Bub(2020{\natexlab{a}})}]{bub2018a}
Bub, J. 
\newblock In defense of a ``single-world'' interpretation of quantum mechanics.
\newblock {\em Stud. Hist. Philos. Mod. Phys.} \textbf{2020}, {\em
  72\/}, 251--255.


\bibitem[{Bub(2020{\natexlab{b}})}]{bubRedux}
Bub, J. 
\newblock `{T}wo {D}ogmas' redux.
\newblock In  {\em Quantum, Probability, Logic:
  The Work and Influence of {I}tamar {P}itowsky\/}; Hemmo, M., Shenker, O., Eds.; 
  Springer: Cham, Switzerland, 2020; pp.\ 199--215.


\bibitem[{Bub(2021)}]{bub2021}
Bub, J. 
\newblock Understanding the {F}rauchiger-{R}enner argument.
\newblock {\em Found. Phys.\/} \textbf{2021}, {\em 51\/}, 36.


\bibitem[{Cuffaro(2010)}]{cuffaro2010}
Cuffaro, M.E. 
\newblock The {K}antian framework of complementarity.
\newblock {\em Stud. Hist. Philos. Mod. Phys.\/} \textbf{2010}, {\em
  41\/}, 309--317.


\bibitem[{Cuffaro(2018)}]{cuffaro2018d}
Cuffaro, M.E. 
\newblock {K}antian and Neo-{K}antian First Principles for Physical and
  Metaphysical Cognition. 2018.
\newblock Available online: \url{philsci-archive.pitt.edu/21625/} (accessed on 29 September 2023).


\bibitem[{Cuffaro(Forthcoming)}]{cuffaroPerovicReview}
Cuffaro, M.E. 
\newblock Review of ``{F}rom data to quanta: {N}iels {B}ohr's vision of
  physics,'' by {S}lobodan {P}erovi\'c.
\newblock {\em Philos. Sci.\/} 
\newblock 2023,
 \emph{forthcoming}.
  \url{https://doi.org/10.1017/psa.2023.108}.


\bibitem[{Bohr(1928{\natexlab{b}})}]{bohr1928}
Bohr, N. 
\newblock The quantum postulate and the recent development of atomic theory.
\newblock {\em Nature\/} \textbf{1928}, {\em 121\/}, 580--590.


\bibitem[{{De Gregorio}(2014)}]{degregorio2014}
{De Gregorio}, A. 
\newblock Bohr's way to defining complementarity.
\newblock {\em Stud. Hist. Philos. Mod. Phys.\/} \textbf{2014}, {\em
  45\/}, 72--82.


\bibitem[{Bohr(1928{\natexlab{a}})}]{bohrToDirac1928}
Bohr, N. 
\newblock Private letter to Paul Dirac, 24 March. 1928.
\newblock Reprinted in Niels Bohr, Collected Works, J{\o}rgen Kalckar
  Ed.; North-Holland/Elsevier:  Amsterdam, The Netherlands,
 1985,  Volume 6, pp.\ 45--46.


\bibitem[{Stein(1994)}]{stein1994}
Stein, H. 
\newblock Some reflections on the structure of our knowledge in physics.
\newblock In {\em Logic,
  Metholodogy and Philosophy of Science {IX}\/}; Prawitz, D., Skyrms, B., Westerstahl, D., Eds.;  Elsevier: Amsterdam, The Netherlands, 
1994; pp.\ 633--655.


\bibitem[{Curiel(2020)}]{curiel2020}
Curiel, E. 
\newblock Schematizing the observer and the epistemic content of theories. \emph{arXiv} \textbf{2020}, 
\newblock {arXiv:1903.02182v3}.


\bibitem[{Stein(1989)}]{stein1989}
Stein, H. 
\newblock Yes, but ... some skeptical remarks on realism and anti-realism.
\newblock {\em Dialectica\/} \textbf{1989}, {\em 43\/}, 47--65.


\bibitem[{Stein(1972)}]{stein1972}
Stein, H. 
\newblock On the conceptual structure of quantum mechanics.
\newblock In {\em Paradigms and Paradoxes: The
  Philosophical Challenge of the Quantum Domain\/};  Colodny, R.G., Ed.;   
  University of Pittsburgh Press: Pittsbugh, PA, USA, 1972; pp.\ 367--438.


\bibitem[{Myrvold(2011)}]{myrvold2011}
  Myrvold, W.C. 
  \newblock Statistical Mechanics and Thermodynamics: A {M}axwellian View
  \newblock \emph{Stud. Hist. Philos. Mod. Phys.} \textbf{2011}, \emph{42\/}, 237--243.

  
\bibitem[{Bogen \& Woodward(1988)}]{bogenwoodward1988}
Bogen, J.; Woodward, J. 
\newblock Saving the phenomena.
\newblock {\em  Philos. Rev.\/} \textbf{1988}, {\em 97}, 303--352.


\bibitem[Newton(1999)]{newton1999}
Newton, I. 
\newblock Mathematical principles of natural philosophy.
\newblock In {\em The {P}rincipia: {A} New Translation and
  Guide\/}; Cohen, I.B., Ed.;   University of
  California Press: Berkely, CA, USA;  Los Angeles, CA, USA, 1999;
 pp.\ 371--946.


\bibitem[{Harper(2011)}]{harper2011}
Harper, W.L.
\newblock {\em Isaac Newton's Scientific Method\/}; 
\newblock Oxford University Press: Oxford, UK, 2011.


\bibitem[{Smith(2002)}]{smith2002}
Smith, G.E. 
\newblock The methodology of the \emph{Principia}.
\newblock In  {\em The Cambridge Companion to
  Newton\/}; Cohen, I.B.,   Smith, G.E., Eds.; Cambridge University Press: Cambridge, UK, 2002; pp.\ 138--173.


\bibitem[{Boole(1862)}]{boole1862}
Boole, G.
\newblock On the theory of probabilities.
\newblock {\em Philos. Trans. R. Soc. Lond.\/}
  \textbf{1862}, {\em 152\/}, {225--252}.


\bibitem[{Pitowsky(1994)}]{pitowsky1994}
Pitowsky, I.
\newblock {G}eorge {B}oole's `conditions of possible experience' and the
  quantum puzzle.
\newblock {\em Br. J. Philos. Sci.} \textbf{1994}, {\em 45\/},
  99--125.


\bibitem[{Pitowsky(1991)}]{pitowsky1991}
Pitowsky, I.
\newblock Correlation polytopes, their geometry and complexity.
\newblock {\em Math. Program. A\/} \textbf{1991}, {\em 50\/}, 395--414.


\bibitem[{Pitowsky(2008)}]{pitowsky2008}
Pitowsky, I. 
\newblock Geometry of quantum correlations.
\newblock {\em Phys. Rev. A\/} \textbf{2008}, {\em 77\/}, 062109.


\bibitem[{Mermin(1981)}]{mermin1981}
Mermin, N.D. 
\newblock Quantum mysteries for everyone.
\newblock {\em J. Philos.\/} 1981, {\em 78\/}, 397--408.


\bibitem[{Von~Neumann(1927)}]{vonNeumann1927b}
Von~Neumann, J. 
\newblock {W}ahrscheinlichkeitstheoretischer {A}ufbau der {Q}uantenmechanik.
\newblock {\em K\"onigliche Ges. Der Wiss. G\"ottingen. Math.-Phys. Klasse. Nachrichten\/} \textbf{1927}, 245--272.


\bibitem[{Janssen(2009)}]{janssen2009}
Janssen, M. 
\newblock Drawing the line between kinematics and dynamics in special
  relativity.
\newblock {\em Stud. Hist. Philos. Mod. Phys.\/} \textbf{2009}, {\em
  40\/}, 26--52.


\bibitem[{Aaronson(2013)}]{aaronson2013}
Aaronson, S. 
\newblock {\em Quantum Computing Since {D}emocritus\/}; 
\newblock Cambridge University Press: New York, NY, USA,  2013.


\bibitem[{Nielsen \& Chuang(2000)}]{nielsenChuang2000}
Nielsen, M.A.;  Chuang, I.L. 
\newblock {\em Quantum Computation and Quantum Information\/}; 
\newblock  {Cambridge University Press}: Cambridge, UK, 2000.


\bibitem[{Wallace(2019)}]{wallace2019}
Wallace, D. 
\newblock On the plurality of quantum theories: Quantum theory as a framework,
  and its implications for the quantum measurement problem.
\newblock In  {\em Realism and the Quantum\/}; French, S.,   Saatsi, J., Eds.; 
  Oxford University Press:  Oxford, UK,  2019; \mbox{pp.\ 78--102.}


\bibitem[{Myrvold(2022)}]{myrvold2022}
Myrvold, W.C. 
\newblock Philosophical Issues Raised by Quantum Theory and its Interpretations
\newblock In \emph{The Oxford Handbook of the History of Quantum Interpretations}; Freire, O., Ed.;  Oxford University Press: Oxford, UK, 2021; pp.\ 53--76.


\bibitem[{Vaidman(2018)}]{vaidman2018}
Vaidman, L.
\newblock Many-worlds interpretation of quantum mechanics.
\newblock In  {\em The {S}tanford Encyclopedia of
  Philosophy\/}, Fall 2018 ed.; Zalta, E.N.,  Ed.; Metaphysics Research Lab, {S}tanford University: Stanford, CA, USA, 2018.


\bibitem[{MacBride(2022)}]{macbrideTruthmakerSEP}
MacBride, F..
\newblock Truthmakers.
\newblock In {\em The {S}tanford Encyclopedia of
  Philosophy\/}, fall 2022 ed.; Zalta,  E.N., Ed.; Metaphysics Research Lab, {S}tanford University: Stanford, CA, USA, 2022.


\bibitem[{Pitowsky(2006)}]{pitowsky2006}
Pitowsky, I. 
\newblock Quantum mechanics as a theory of probability.
\newblock In  {\em Physical Theory and Its
  Interpretation\/}; Demopoulos, W., Pitowsky, I., Eds.;  Springer:  Dordrecht, The Netherlands, 2006; pp.\ 213--240.


\bibitem[{Bacciagaluppi(2020)}]{bacciagaluppi2020}
Bacciagaluppi, G. 
\newblock The Role of Decoherence in Quantum Mechanics
\newblock In {\em The {S}tanford Encyclopedia of
  Philosophy\/}, fall 2020 ed.; Zalta,  E.N., Ed.; Metaphysics Research Lab, {S}tanford University: Stanford, CA, USA, 2022.


\bibitem[{Crull(2022)}]{crull2022}
Crull, E. 
\newblock The Philosophical Significance of Decoherence. 
\newblock In \emph{Oxford Research Encyclopedia of Physics}; Oxford University Press: Oxford, 2022
\newblock  \url{https://doi.org/10.1093/acrefore/9780190871994.013.78}.


\bibitem[{Frauchiger \& Renner(2018)}]{frauchigerRenner2018}
Frauchiger, D.; Renner, R.
\newblock Quantum theory cannot consistently describe the use of itself.
\newblock {\em Nat. Commun.\/} \textbf{2018}, {\em 9\/}, 3711.


\bibitem[{Dascal(2020)}]{dascal2020}
Dascal, M. 
\newblock What's left for the neo-{C}openhagen theorist?
\newblock {\em Stud. Hist. Philos. Mod. Phys.\/} \textbf{2020}, {\em
  72\/}, 310--321.


\bibitem[{Felline(2020)}]{felline2020a}
Felline, L. 
\newblock The measurement problem and two dogmas about quantum mechanics.
\newblock In  {\em Quantum, Probability,
  Logic\/}; Hemmo, M.,  Shenker, O.,  Eds.; Springer: Cham, Switzerland,   2020; pp.\ 285--304.


\bibitem[Bell(1987)]{bell1973}
Bell, J.S. 
\newblock Subject and object.
\newblock In {\em Speakable and Unspeakable in Quantum Mechanics\/};  Cambridge University Press: Cambridge, UK, 1987;  pp.\ 40--44.


\bibitem[{Acuna(2021)}]{acuna2021}
Acu\~na, P.
\newblock {V}on {N}eumann's theorem revisited.
\newblock {\em Found. Phys.\/} \textbf{2021}, {\em 51:73\/}, 1--29.


\bibitem[{Bub(2010)}]{bub2010b}
Bub, J. 
\newblock {V}on {N}eumann's `no hidden variables' proof: A re-appraisal.
\newblock {\em Found.  Phys.\/} \textbf{2010}, {\em 40\/}, 1333--1340.


\bibitem[{Dieks(2017)}]{dieks2017}
Dieks, D. 
\newblock {V}on {N}eumann's impossibility proof: Mathematics in the service of
  rhetorics.
\newblock {\em Stud. Hist. Philos. Mod. Phys.\/} \textbf{2017}, {\em
  60\/}, 136--148.


\bibitem[{Ghirardi(2018)}]{ghirardi2018}
Ghirardi, G. 
\newblock Collapse theories.
\newblock In {\em The {S}tanford Encyclopedia of
  Philosophy\/}, {Fall} 2018
  ed.; Zalta, E.N., Ed.; Metaphysics Research Lab, {S}tanford University:  Stanford, CA, USA, 2018.


\bibitem[{Goldstein(2009)}]{goldstein2006}
Goldstein, S. 
\newblock Bohmian mechanics.
\newblock In {\em The {S}tanford Encyclopedia of
  Philosophy\/}, Spring 2009  ed.;  Zalta, E.N., Ed.;  Metaphysics Research Lab, {S}tanford University: Stanford, CA, USA, 2009.


\bibitem[{Adlam(2022)}]{adlam2022}
Adlam, E. 
\newblock Does science need intersubjectivity? {T}he problem of confirmation in
  orthodox interpretations of quantum mechanics.
\newblock {\em Synthese\/} \textbf{2022}, {\em 200\/}, 522.


\bibitem[{Bohr(1958)}]{bohr1958}
Bohr, N. 
\newblock Quantum physics and philosophy.
\newblock In {\em Philosophy in the Mid-Century: A
  Survey\/};  Klibansky, R., Ed.; La Nuova Italia Editrice: Firenze, Italy, 1958; pp.\ 308--314.


\bibitem[{Burgess(1928)}]{burgess1928}
  Burgess, G. 
  \newblock The International Temperature Scale.
  \newblock \emph{J. Res. Natl. Bur. Stand.} \textbf{1928}.


\bibitem[{Chang(2021)}]{chang2021}
  Chang, H. 
  \newblock Operationalism
  \newblock In {\em The {S}tanford Encyclopedia of
  Philosophy\/}, Fall 2021  ed.;  Zalta, E.N., Ed.;  Metaphysics Research Lab, {S}tanford University: Stanford, CA, USA, 2021.


\bibitem[{Myrvold(2010)}]{myrvold2010}
  Myrvold, W.C. 
  \newblock From Physics to Information Theory and Back.
  \newblock In  \emph{Philosophy of Quantum Information and Entanglement};  Bokulich, A.,  Jaegger, G., Eds.; Cambridge University Press: Camrbidge, UK, 2010; pp.\ 181--207.


\bibitem[{Myrvold(2021)}]{myrvold2021}
Myrvold, W.C. 
\newblock {\em Beyond Chance and Credence: A Theory of Hybrid Probabilities\/};
\newblock Oxford University Press: Oxford, UK, 2021.


\bibitem[{Hansen \& Wolf(2019)}]{hansenWolfFeatureNotBug}
Hansen, A.; Wolf, S. 
\newblock Contextuality: It's a feature, not a bug. \emph{arXiv} \textbf{2019}, 
\newblock {arXiv:1902.02088}.


\bibitem[{Hughes(1989)}]{hughes1989}
Hughes, R.I.G. 
\newblock \emph{The Structure and Interpretation of Quantum Mechanics};
\newblock  Harvard University Press: Cambridge, MA, USA, 1989.


\bibitem[{Cuffaro \& Hartmann(2023)}]{cuffaroHartmannOpenSystemsView}
Cuffaro, M.E.; Hartmann, S. 
\newblock The open systems view. \emph{arXiv} \textbf{2023},
\newblock {arXiv:2112.11095v2}.


\bibitem[{Einstein(1948)}]{einstein1948}
Einstein, A. 
\newblock Quanten-mechanik und {W}irklichkeit.
\newblock {\em Dialectica\/} \textbf{1948}, {\em 2\/}, 320--324.


\bibitem[{Howard(1985)}]{howard1985}
Howard, D. 
\newblock Einstein on locality and separability.
\newblock {\em Stud. Hist. Philos. Sci.\/} \textbf{1985}, {\em 16\/},
  171--201.


\bibitem[{Ram\'irez(2020)}]{ramirez2020}
Ram\'irez, S. 
\newblock Separating {E}instein's separability.
\newblock {\em Stud. Hist. Philos. Mod. Phys.\/} \textbf{2020}, {\em 72\/}, 138--149.


\bibitem[{Wallace(2022)}]{wallaceIsolatedSystemsI}
Wallace, D.
\newblock Isolated systems and their symmetries, part {I}: General framework
  and particle-mechanics examples.
\newblock {\em Stud. Hist. Philos. Sci.\/} \textbf{2022}, {\em 92\/},
  239--248.


\bibitem[{Lehner(2014)}]{lehner2014}
Lehner, C. 
\newblock {E}instein's realism and his critique of quantum mechanics.
\newblock In {\em The {C}ambridge Companion to
  {E}instein\/}; Janssen, M., Lehner, C., Eds.; Cambridge University Press: Cambridge, UK, 2014; pp.\ 306--352.


\bibitem[{Cuffaro(2020)}]{cuffaroInfCausality}
Cuffaro, M.E. 
\newblock Information causality, the {T}sirelson bound, and the `being-thus' of
  things.
\newblock {\em Stud. Hist. Philos. Mod. Phys.\/} \textbf{2020}, {\em
  72\/}, 266--277.


\bibitem[{d'Espagnat(1966)}]{despagnat1966}
d'Espagnat, B. 
\newblock An elementary note about `mixtures'.
\newblock In  {\em Preludes in
  Theoretical Physics\/}; de~Shalit, A., Feshback, H.,   van Hove, L., Eds.; North Holland Wiley: Amsterdam, The Netherlands, 1966; pp.\ 185--191.


\bibitem[{d'Espagnat(1976)}]{despagnat1976}
d'Espagnat, B. 
\newblock {\em Conceptual Foundations of Quantum Mechanics (2nd edition)\/}.
\newblock W.\ A.\ Benjamin: Menlo Park, CA, USA, 1971. 


\bibitem[{Wallace \& Timpson(2010)}]{wallace2010a}
Wallace, D.; Timpson, C.G. 
\newblock Quantum mechanics on spacetime {I}: Spacetime state realism.
\newblock {\em Br. J. Philos. Sci.\/} \textbf{2010}, {\em 61\/},
  697--727.


\bibitem[{Curiel(2014)}]{curiel2014}
Curiel, E. 
\newblock Classical mechanics is {L}agrangian; it is not {H}amiltonian.
\newblock {\em  Br. J. Philos. Sci.\/} \textbf{2014}, {\em 65\/},
  269--321.


\bibitem[{Popper(1959)}]{popper1959}
Popper, K.R. 
\newblock The propensity interpretation of probability
\newblock {\em Br. J. Philos. Sci.\/} \textbf{1959}, {\em 10\/}, 25--42.


\bibitem[{Bub(2022)}]{bubForewordToRaffles}
Bub, J. 
\newblock Foreword.
\newblock In {\em Understanding Quantum Raffles: Quantum Mechanics on an
  Informational Approach: Structure and Interpretation\/};  
\newblock Springer: Cham, Switzerland,  
\newblock  2022; pp. viii--xiv.

\end{thebibliography}
\end{document}